\documentclass[iop,apjl,numberedappendix]{emulateapj-rtx4}
\usepackage{graphicx}
\usepackage{mathrsfs}
\usepackage[intlimits,centertags]{amsmath}
\usepackage{amssymb,amsfonts}
\usepackage[pdftex]{hyperref}
\usepackage[x11names]{xcolor}
\hypersetup{
pdfstartview={FitH},
colorlinks=true,
bookmarksopen=false,
bookmarksnumbered=false,
bookmarksopenlevel=0,
linkcolor=Blue1!70!black,
citecolor=Blue1!70!black,
urlcolor=Blue1!70!black
}

\begin{document}

\title{A search for neutrinos from fast radio bursts with IceCube}


\author{Samuel Fahey}
\author{Ali Kheirandish}
\author{Justin Vandenbroucke\altaffilmark{1}}
\author{Donglian Xu}
\affil{Wisconsin IceCube Particle Astrophysics Center and Department of Physics,\\ University of Wisconsin, Madison, WI 53706, USA}
\altaffiltext{1}{justin.vandenbroucke@wisc.edu}

\begin{abstract}
We present a search for neutrinos in coincidence in time and direction with four fast radio bursts (FRBs) detected by the Parkes and Green Bank radio telescopes during the first year of operation of the complete IceCube Neutrino Observatory (May 2011 through May 2012).  The neutrino sample consists of 138,322 muon neutrino candidate events, which are dominated by atmospheric neutrinos and atmospheric muons but also contain an astrophysical neutrino component.  Considering only neutrinos detected on the same day as each FRB, zero IceCube events were found to be compatible with the FRB directions within the estimated 99\% error radius of the neutrino directions.  Based on the non-detection, we present the first upper limits on the neutrino fluence from fast radio bursts.
\end{abstract}


\maketitle

\section{Introduction}

Fast radio bursts (FRBs) are a new class of astrophysical radio transients of very short (few millisecond) duration.  The first was discovered in a 2007 analysis of archival data from the Parkes telescope~\citep{Lorimer2007}.  A total of 23 unique burst directions have now been detected by five different telescopes (Parkes, Arecibo, Green Bank, UTMOST, and ASKAP;~\citet{FRBCAT}).  One (and only one) of these directions has been found to repeat, producing at least 17 bursts at different times~\citep{repeating, repeatingAgain}.  The first claimed host identification and redshift~\citep{host} for an FRB was later shown to be an active galactic nucleus~\citep{noHost}.  However, after precise localization of the repeating burst by the VLA, an optical host galaxy was found~\citep{localization}, and its redshift was determined to be 0.19~\citep{redshift}.  Given their rate of detection by radio surveys performed with relatively low exposure time and field of view, the rate of FRBs across the full (4$\pi$) sky is estimated to be several thousand per day~\citep{Champion}, about 10\% of the core-collapse supernova rate~\citep{Murase}.

The origin and emission mechanism of these bursts is unknown.  Models have proliferated and include the birth of black holes from supramassive neutron stars~\citep{Falcke2014} and giant flares from magnetars~\citep{magnetar}.  Their large dispersion measures indicate an extragalactic origin, but they could also originate in Galactic sources enshrouded in dense plasma~\citep{Loeb2014}.  Only one burst has been proven to repeat, and the same burst is the only one proven to be extragalactic.  While the repetition rules out a cataclysmic model for that source, other bursts could be produced in cataclysmic scenarios.  While leptonic emission is the default assumption in most models, hadronic emission mechanisms or association with hadronic emission regions are also possible, with implications for cosmic rays and neutrino emission~\citep{Li2013}.

A 15-50~keV gamma-ray transient of $\sim$300~s duration, coincident with FRB~131104 with a statistical significance of $\sim$3~sigma, was reported by~\cite{DeLaunay}.  No other afterglow or counterpart has been detected.  If this is a genuine counterpart, the gamma-ray fluence is $\sim$6 orders of magnitude greater than the radio fluence, raising the energy budget for modeling the emission and for additional counterparts.

In addition to energy budget considerations, there are two additional constraints in modeling neutrino emission from FRBs: (1) the neutrino emission region must be dense enough in target matter or radiation to produce neutrinos, but not dense enough to absorb radio emission if it is produced in the same region; (2) it is difficult to cool hadrons quickly and thereby produce neutrino emission on short time scales.  Nevertheless, target radiation fields and baryons are expected in the environment surrounding many possible FRB progenitors~\citep{Murase}.

Because of their very short duration, prompt counterparts are most likely to be detected either by coordinated observation campaigns or serendipitously, in the latter case most likely by wide-field instruments.  Because there is still so little known about the nature of fast radio bursts, it is essential to perform model-independent searches using a variety of wide-field instruments spanning multiple wavelengths and messengers.



IceCube is a cubic-kilometer neutrino detector located at the geographic South Pole.  It consists of an array of 5160 digital optical modules encompassing a gigaton of ice as the active volume~\citep{detector}. With sensitivity to all neutrino flavors over the full sky including both hemispheres, the IceCube detector enables a wide range of science~\citep{Achterberg2006155}.

IceCube has discovered a diffuse astrophysical neutrino flux between several TeV and several PeV~\citep{IceCubePeV, HESE1, HESE2, mese, muonNeutrinos, globalFit, numu6year}.  Many of the neutrino events originate far from the Galactic plane and are therefore likely to be extragalactic.  Although the diffuse flux is detected with high statistical significance in multiple distinct detection channels, no evidence for discrete sources has been found, either in searches for clustering of the neutrinos or in cross-correlation with catalogs of source candidates~\citep{0004-637X-796-2-109}.  The origin of the astrophysical neutrinos remains unknown.  The majority of the astrophysical flux is not produced by gamma-ray bursts~\citep{NatureGRB,GRB2015,GRB2017} or star-forming galaxies~\citep{SFG}.

\section{Neutrino Sample}\label{evt_selec}


The event sample used in this analysis is part of a multi-year data set optimized to search for point sources of astrophysical neutrinos.  The event selection is described in detail in~\cite{0004-637X-796-2-109}.  The first year of events and accompanying details (including the effective area of the event selection as a function of energy and declination) were recently released~\citep{dataRelease}.  For each event, the data release includes the time of the event truncated to integer Modified Julian Day (MJD), the best-fit energy and direction, and an estimate of the direction uncertainty (50\% containment radius).


\begin{figure}[b]
\centering
\includegraphics[width=0.5\textwidth]{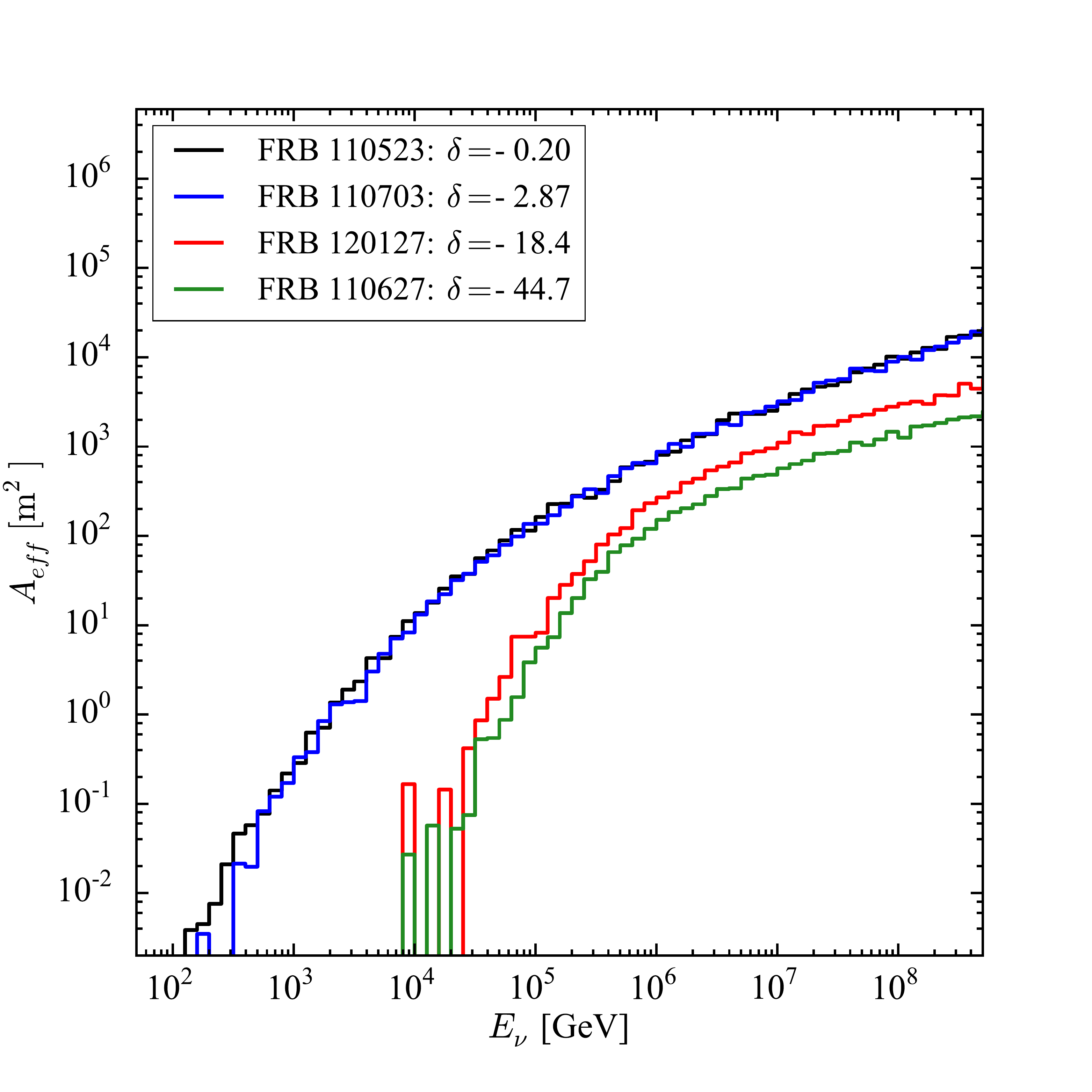} 
\caption{Neutrino effective area as a function of energy for the event selection used in this analysis, in the direction of each FRB.  The declination $\delta$ of each FRB is given (in degrees) in the legend.  The effective area in the southern sky is less than that near the celestial equator due to tighter cuts used to reduce the atmospheric muon background in the southern sky.}
\label{Aeff}
\end{figure}

\begin{figure}[b]
\centering
\includegraphics[width=0.5\textwidth]{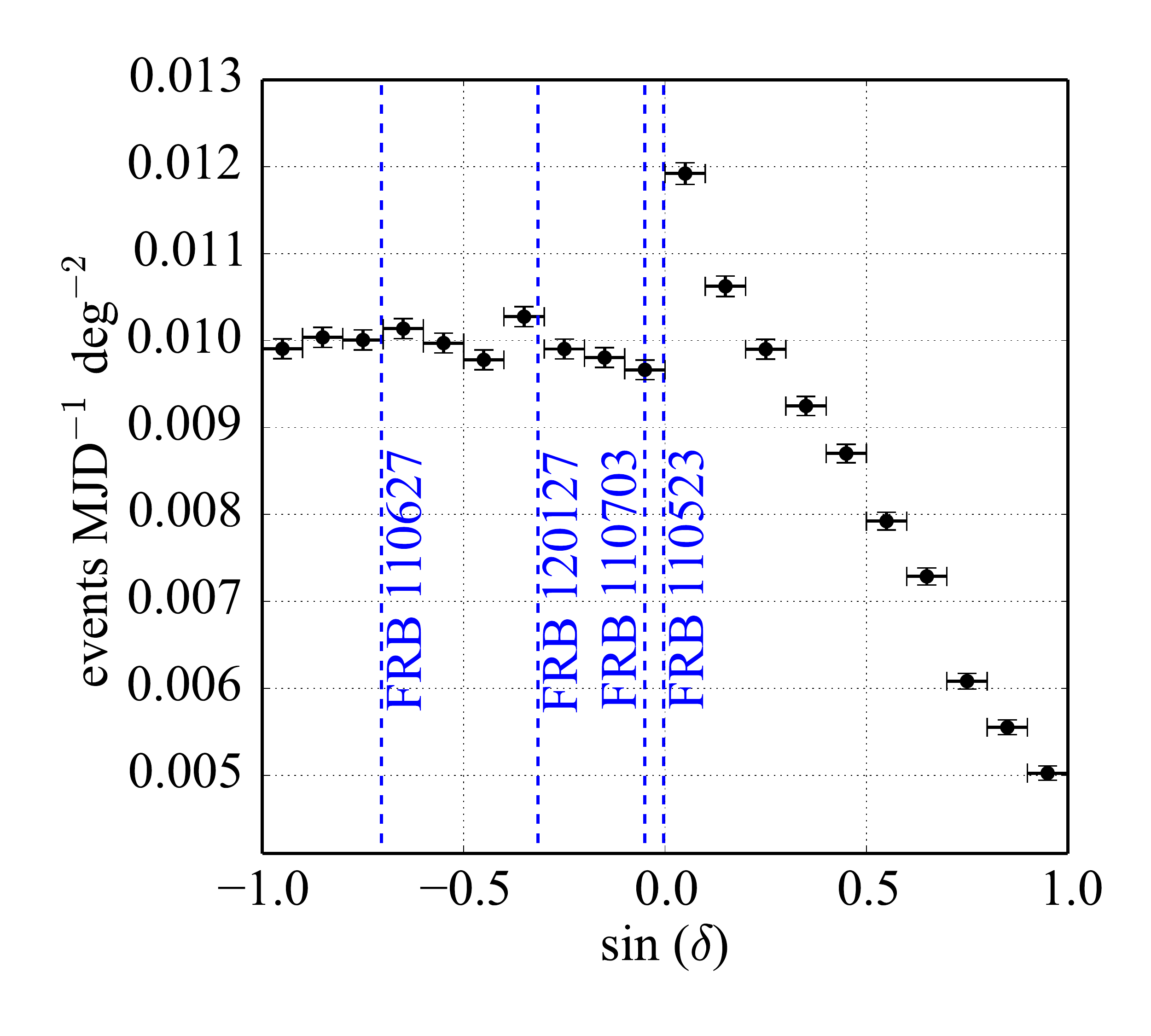} 
\caption{Event rate in the IceCube data sample as a function of declination, averaged over right ascension within each declination band. The declination of each FRB is shown for reference.  The rate is normalized per calendar day between MJD 55694 and 56062 (369 days), not per day of livetime.}
\label{evt_dist}
\end{figure}

The data set includes a total of 138,322 events from 333 days of livetime spanning May 2011 to May 2012 (MJD 55694 through 56062), with a roughly equal number of events from the northern and southern hemispheres. The data reduction and event reconstruction procedures are detailed in \citet{0004-637X-796-2-109}. Events with declination greater than $-5^\circ$ are considered ``up-going'' (northern hemisphere) events and are predominantly atmospheric neutrinos. 
``Down-going'' (southern hemisphere) events reconstructed to originate from declination less than $-5^
\circ$ are dominated by cosmic-ray-induced atmospheric muons and high-energy muon bundles (multiple muons produced in the same extensive air shower).

As discussed in \citet{0004-637X-796-2-109}, the event selection was performed separately for the northern and southern hemispheres with boosted decision trees. In the up-going region, the ice and the Earth act as a shield for atmospheric muons, so a high-purity neutrino sample with a wide energy range and low energy threshold is obtained. In the down-going region, high-energy neutrinos are also retained, but a high-purity  neutrino sample cannot be as easily achieved due to the atmospheric muons. In order to bring the atmospheric muon contamination under control, a higher energy threshold was applied in the southern sky.

Figure~\ref{Aeff} shows the muon neutrino effective area of the IceCube event selection as a function of energy in the direction of each FRB.  At every energy the effective area is smaller in the southern sky than near the celestial equator due to the tight cuts necessary to reduce the cosmic-ray muon background in the southern sky.  In the southern sky, fluctuations are visible in the effective area curves near the energy threshold (
$\sim$20~TeV), likely due to statistical fluctuations in the Monte Carlo used to calculate the curves.

The rate of detected events in the sample varies from day to day due to natural causes such as seasonal variation in the production of atmosphere neutrinos and muons~\citep{seasonal} and due to detector effects  such as downtime.  We estimated the size of possible downtime effects from the number of IceCube events detected on the day of each FRB.  The event counts are (in time order of FRB occurrence) 423, 395, 342, and 465. Because the event count on each day is within $\sim$20\% of the average count per day in the full sample (375), detector deadtime was likely not substantial on any of the FRB days.

Figure~\ref{evt_dist} shows the event rate in this sample as a function of declination, averaged over right ascension and time during the year.  Because of the higher energy threshold applied in the southern hemisphere to counteract the high atmospheric muon rate, the event rate varies by only a factor of $\sim$2 across the sky.  The average rate is 0.009 events per square degree (roughly the area of the point-spread function) per day.  Detection of a single event compatible with the direction of an FRB and detected on the same day as the FRB would therefore be interesting.

\section{Coincidence search}\label{ana}

\begin{figure}[th]
\includegraphics[width=0.5\textwidth]{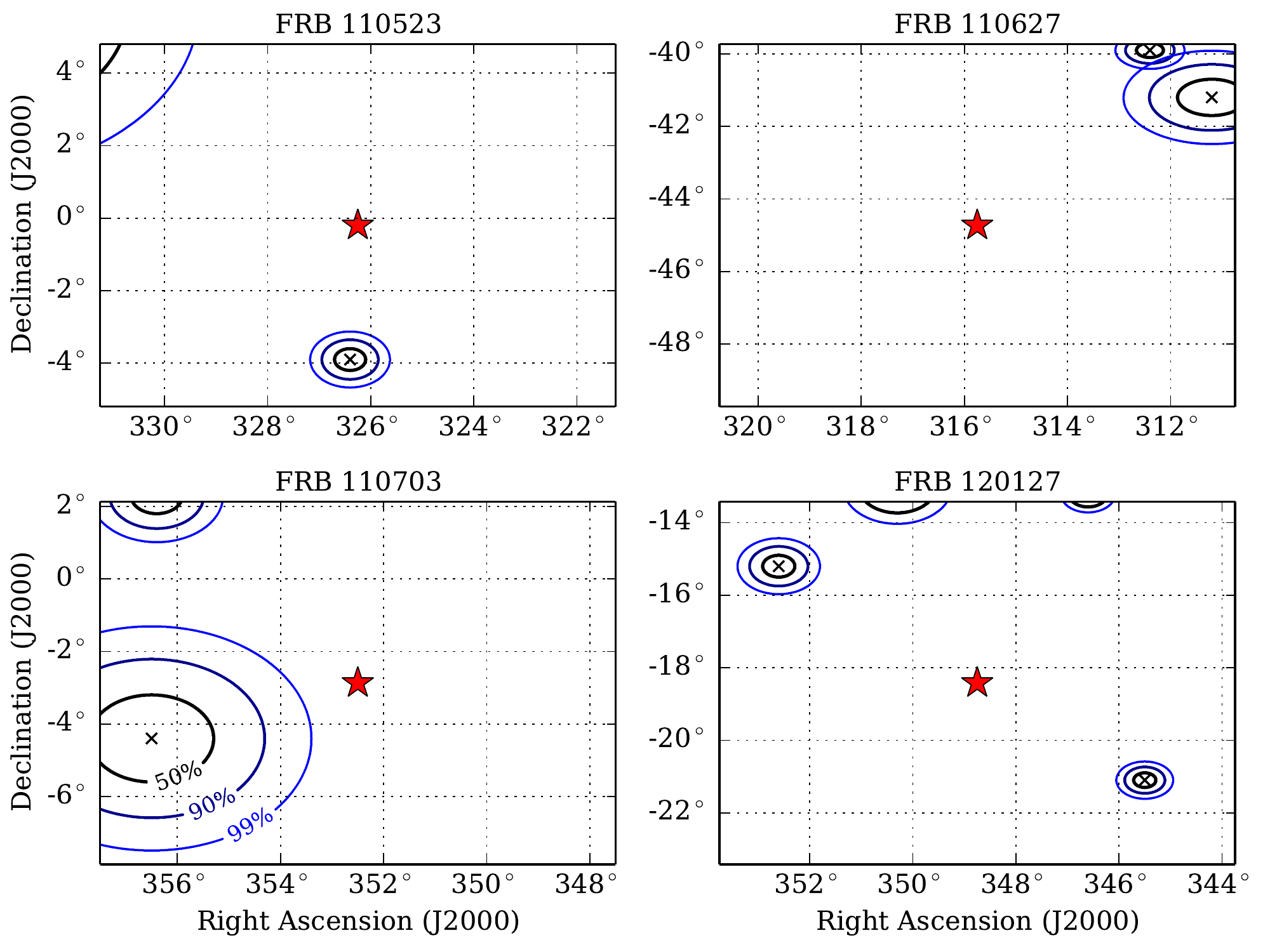} 
\caption{The region of interest centered on each FRB ($\star$) in this sample is shown in equatorial coordinates in Cartesian projection. The best-fit direction of each IceCube event is indicated with an $\times$.  The 50\%-containment circle for each event is shown, as is an estimate of the 90\%- and 99\%-containment circles under the approximation that the point spread function is a radially symmetric two-dimensional Gaussian distribution.}
\label{frb_evt}
\end{figure}



Four FRBs have been detected in the year spanned by this IceCube event sample: FRB 110523~\citep{GBT}, FRB 110627, FRB 110703, and FRB 120127~\citep{Thornton2013}. Two are near the celestial equator and two are well south of it.  Because the IceCube event times are truncated to integer MJD, temporal coincidence with these FRBs can only be tested on the one-day time scale.  For each FRB, the radio burst detection time was truncated to integer MJD and the angular distance to each IceCube event on the same day was calculated. The localization error of each FRB is $\sim$0.2$^{\circ}$ or better~\citep{Thornton2013, GBT}, negligible in this analysis.

We assume for this search that the point-spread function for each event can be approximated by a radially symmetric two-dimensional Gaussian. Under this assumption, the radius of the 90\% and 99\% error circles can be determined from the 50\% error circle by multiplying by a factor of 1.82 and 2.58, respectively. Figure~\ref{frb_evt} shows these error circles for coincident (on the same truncated MJD as the FRB) events near each of the FRBs.  The nearest (relative to its error circle) coincident event is separated by 4.27$^\circ$ from FRB 110703 on MJD 55745, with a 50\% angular error of 1.2$^\circ$. 



\section{Results and Conclusion}\label{res}

Because there is no IceCube event consistent with the time and direction of any of the four FRBs analyzed, we proceed to constrain the neutrino emission associated with each burst.   Using the Poisson distribution, we construct a 90\% confidence level upper limit on the neutrino fluence by finding the fluence that would produce on average 2.3 detected neutrinos.
 
The expected number of muon neutrinos detected from a source at zenith angle $\theta$ is

\begin{eqnarray}\label{nuevents}
N_{\nu_\mu+\overline{\nu}_\mu} = \int \phi(E_\nu) \,A_{eff}(E_\nu,\theta)\,dE_\nu\,dt ,
\label{Nevent}
\end{eqnarray}

\noindent where $\phi(E_\nu)$ is the neutrino flux at earth and $A_{eff}$ is the IceCube effective area as a function of neutrino energy and zenith angle.  We used the effective area corresponding to the event selection and selected $A_{eff}(E_{\nu})$ for each FRB based on its declination (Figure~\ref{Aeff}). In order to constrain the neutrino flux, we assume the flux to be a power law given by

\begin{eqnarray}
\phi(E_\nu) = \phi_0 \big( \frac{E_\nu}{E_0} \big)^{-\gamma}. 
\end{eqnarray}

\begin{figure}[t]
\centering
\includegraphics[width=0.5\textwidth]{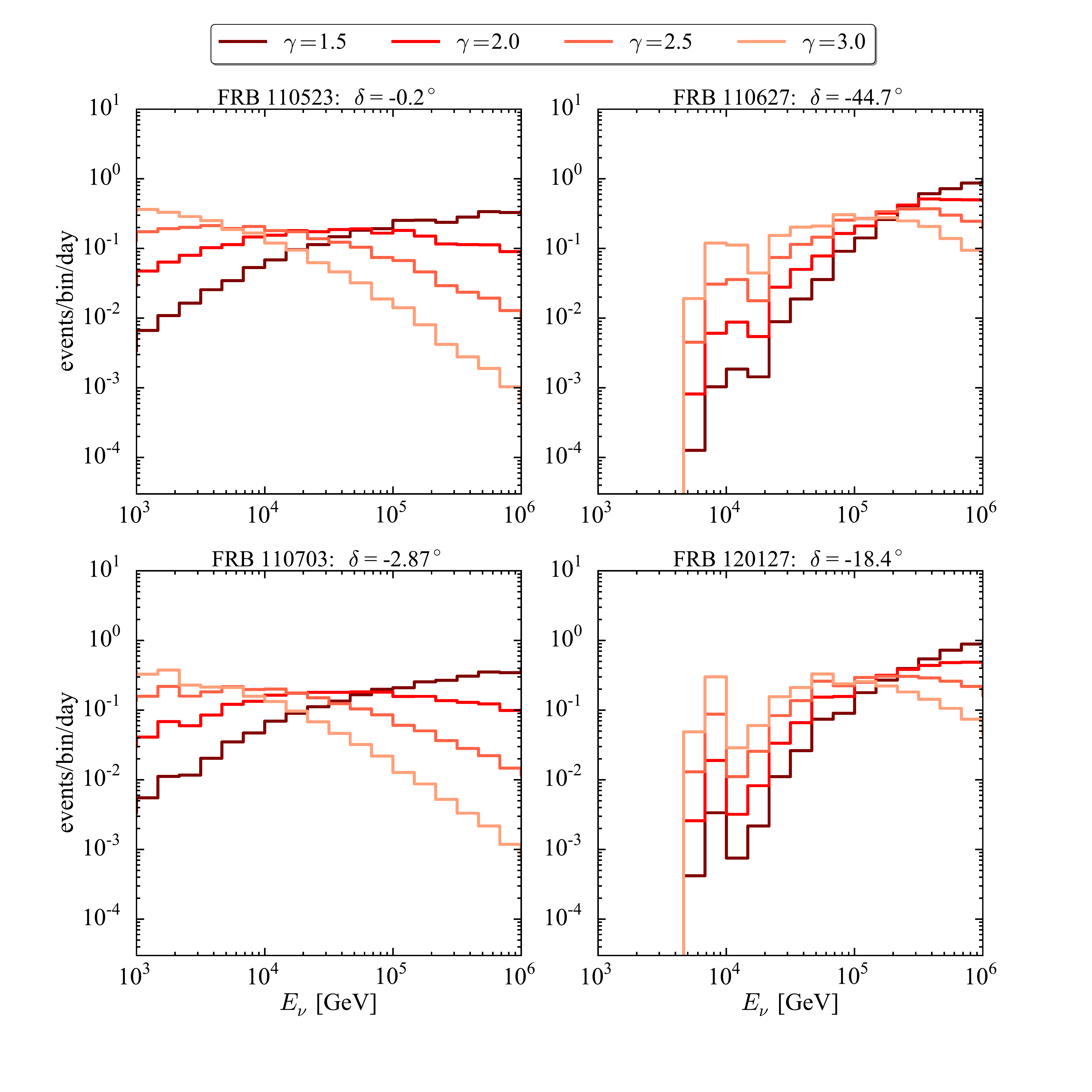} 
\caption{Energy distribution of events that would be detected if the neutrino flux saturated our upper limits.  Each curve is determined by multiplying the power-law spectral model by the detector effective area and normalizing so that the integral is 2.3 events (the 90\% confidence level upper limit on the event rate given that zero events were detected).  Several power law indices ($\gamma$) were tested. \label{dNdE}}
\end{figure}

We set the normalization energy, $E_0$, to 100 TeV and consider four different spectral indices ranging from 1.5 to 3. To calculate the expected number of events we perform the integral in Equation~\ref{nuevents} from 1 TeV to 1 PeV in neutrino energy.

Figure~\ref{dNdE} shows for each burst the distribution of event energies that IceCube would detect for various power-law neutrino spectra.  The shape of each curve is determined by multiplying the flux by the effective area, and each curve is normalized to 2.3 total events, i.e. to the 90\% confidence level upper limit on the expected number of events detected from the burst. As the figure shows, the tightest limits arise from the FRBs found near the celestial equator. This is a result of IceCube's effective area peaking in this direction.

For the two bursts well south of the celestial equator, the effective area curves at these declinations have large fluctuations near $\sim$10~TeV, likely due to statistical uncertainty close to the energy threshold in the Monte Carlo used to determine the effective area.  This is the cause of the fluctuations seen at $\sim$10~TeV in the right two panels of Figure~\ref{dNdE}.

\begin{figure}[h]
\centering
\includegraphics[width=0.5\textwidth]{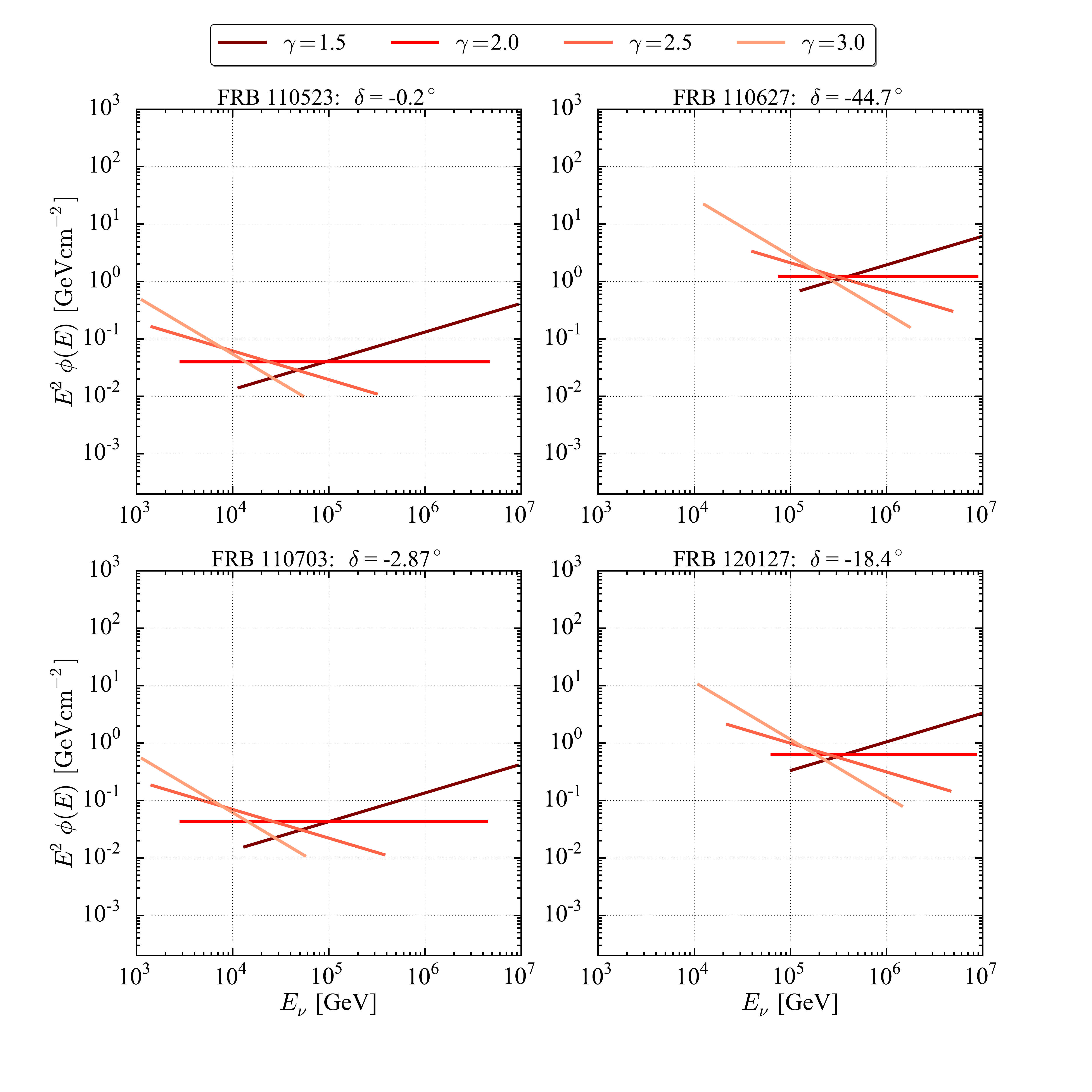} 
\caption{Upper limits ($90\%$ confidence level) on the time-integrated neutrino flux from each FRB, assuming a power-law neutrino spectrum with index $\gamma$.  Each upper limit is drawn over the energy range that contains the central 90\% of events that would be detected (the central 90\% of the distribution shown in Figure~\ref{dNdE}), i.e. the core energy range within which the analysis is sensitive for each burst and spectral model. \label{fluxes}}
\end{figure}

Figure~\ref{fluxes} shows the corresponding time-integrated flux upper limits for several assumed spectral models for each FRB.

The neutrino fluence (time-integrated energy flux) is 

\begin{eqnarray}
f = \int_{E_{min}}^{E_{max}} E\,\phi(E)\,dE\,dt,
\end{eqnarray}

where $E_{min}$ = 1~TeV and $E_{max}$ = 1~PeV.  Table~\ref{tab:1} shows the neutrino fluence upper limit for each burst for $\gamma = 2.0$.

In the future, a more sensitive search can be performed for high-energy neutrinos from these and additional FRBs both by analyzing subsequent years of IceCube data and by using a looser event selection with greater effective area and greater background rate but on shorter time scales (extending from the one-day scale studied here over a range of scales all the way down to the intrinsic $\sim$10~ms FRB duration), similar to the strategy used for gamma-ray-burst neutrino searches~\citep{NatureGRB,GRB2015,GRB2017}. Furthermore, an IceCube search for lower-energy (MeV scale) neutrinos can be performed using an analysis strategy similar to that used for nearby supernovae~\citep{Abbasi:2011ss}.

\begin{figure}[h]
\centering
\includegraphics[width=0.5\textwidth]{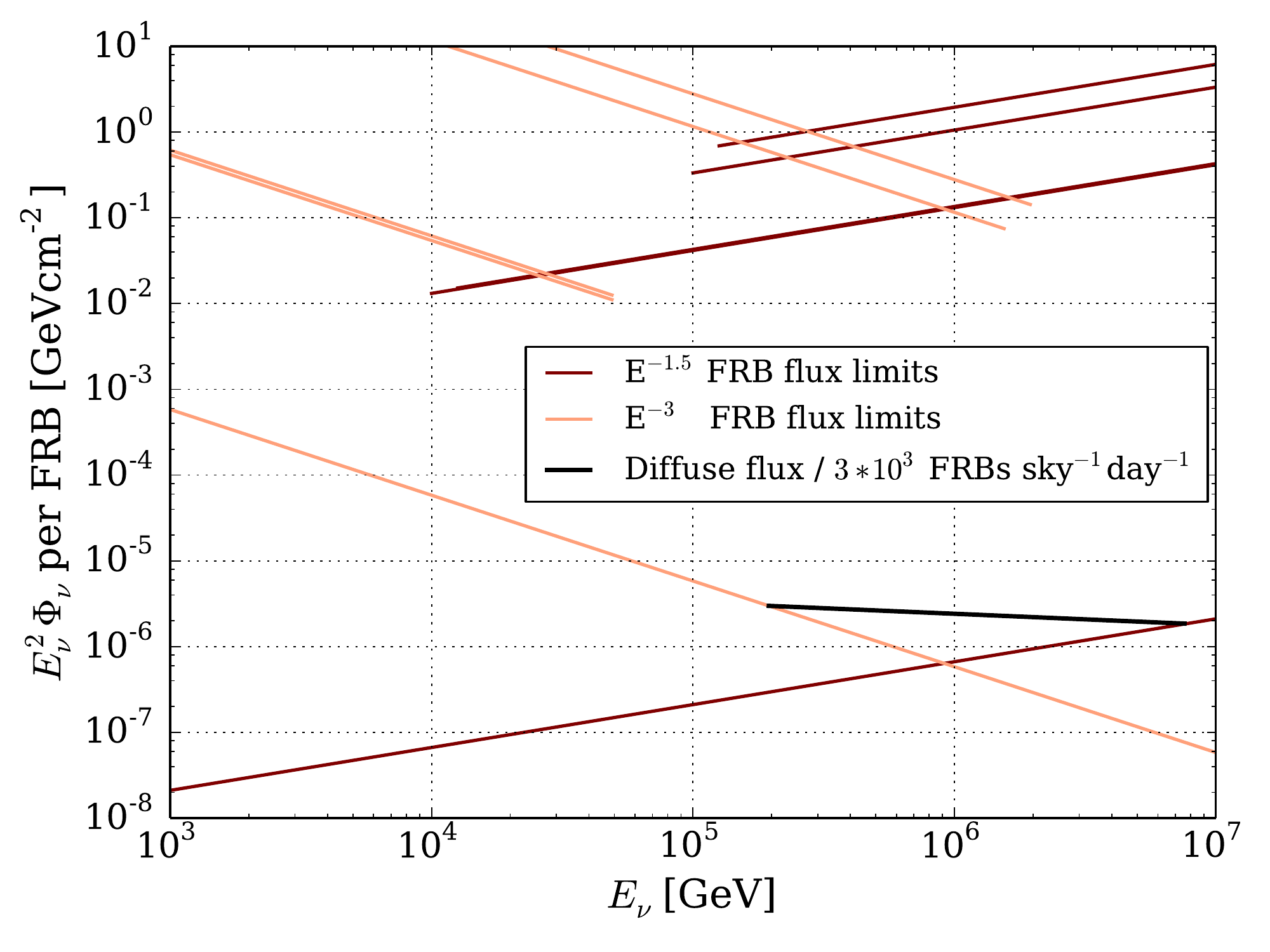} 
\caption{The neutrino fluence upper limits for spectral indices 1.5 and 3.0 from Figure~\ref{fluxes} (upper curves) are compared to limits for the same indices derived by the constraint that FRBs not over-produce the diffuse astrophysical $\nu_\mu$ flux observed by IceCube at any particular neutrino energy, assuming equal neutrino flux from each of $3 \times 10^3$ FRBs per sky per day.  The simple limits derived from the total diffuse emission are currently more constraining than the limits set by the dedicated FRB search.  Stacking many FRBs will enable even stronger constraints.\label{diffuse_phi}}
\end{figure}

Using a complementary approach to derive an upper limit on the neutrino fluence per average FRB, we divide the all-sky diffuse neutrino flux by the all-sky FRB rate.  The result is shown in Figure~\ref{diffuse_phi} and yields a more stringent upper limit than the per-burst analysis.  Note that these two sets of upper limits test different hypotheses.  The per-burst analysis tests whether there is significant neutrino emission from any particular burst, while the calculation using the diffuse neutrino flux treats all FRBs as a single homogeneous population.  That is, if all FRBs across the sky (several thousand per day, the vast majority of which are not radio detected) were to produce a neutrino fluence saturating our per-burst upper limits, then they would produce a total diffuse astrophysical neutrino flux greater than that measured by IceCube.

Additional FRB observations are needed to determine whether there are sub-classes (e.g. repeating and non-repeating, or extragalactic and Galactic) of FRBs.  Based on other astrophysical transients such as supernovae and gamma-ray bursts, sub-classes would not be surprising.  The discovery of FRB sub-classes would bridge the gap between the two approaches by enabling searches for neutrino emission from individual sub-classes each with a lower rate than the total. It may be possible in the future to detect neutrino emission from a particular sub-class of FRBs.



In the absence of FRB sub-classes, stacking many FRBs in the future will enable even tighter constraints than those derived from the diffuse astrophysical neutrino flux.  Interferometric arrays expected to begin operation in 2017 could detect 30 or more FRBs per day, or $\sim$10$^4$ per year~\citep{futureRates}.  For short time windows approaching the radio emission duration ($\sim$10~ms), an IceCube search stacking as many as $10^4$ to $10^5$ bursts will have a total background expectation of $10^{-4}$ to $10^{-3}$ events, and the constraints will become tighter than those derived directly from the diffuse neutrino flux.

Detection of a neutrino FRB signal in this analysis would have indicated a hadronic process, providing a strong constraint on FRB origins and emission mechanisms.  It would have also constituted the first evidence for a high-energy astrophysical neutrino source.  In the absence of a detection, we have calculated upper limits on each burst as well as a comparison to the total diffuse astrophysical neutrino flux.  The expected onslaught of FRB detections in the near future will enable sensitive searches for neutrino emission by stacking thousands of bursts.  We encourage theoretical work to provide quantitative predictions that can be tested by these searches.


\begin{table*}[]
  \begin{center}
  \caption{Characteristics of each fast radio burst (right ascension, declination, time, radio fluence, and telescope) and of the nearest IceCube event detected on that day (angular distance from FRB, error radius).  The final column gives the 90\% confidence level upper limit on the neutrino fluence from the burst assuming the neutrino spectrum is a power law with index 2.0.}
  \begin{tabular}{lcrcccccc}
	\hline
	\hline
	FRB & R.A. & Dec. &  FRB MJD & Radio fluence (GeV cm$^{-2}$) & Telescope & $\Delta\Psi_{\nu- \text{FRB}}$ & $\nu$ error~(50\%) & $f^{90\%}$ (GeV cm$^{-2}$) \\
	\hline
    110523 & 21h45$^\prime$ & -00$^{\circ}$12$^\prime$ & 55704.63 & $2.37\times10^{-15}$ & Green Bank & 3.70$^{\circ}$ & 0.3$^{\circ}$ & 0.184\\
    \hline
    110627 & 21h03$^\prime$	& -44$^{\circ}$44$^\prime$ & 55739.90 & $1.75\times10^{-15}$ & Parkes & 4.85$^{\circ}$ & 0.5$^{\circ}$ & 4.84 \\
	\hline
    110703 & 23h30$^\prime$ & -02$^{\circ}$52$^\prime$ & 	55745.79 & $4.49\times10^{-15}$ & Parkes & 4.27$^{\circ}$ & 1.2$^{\circ}$ & 0.184\\
    \hline
    120127 & 23h15$^\prime$ & -18$^{\circ}$25$^\prime$ & 55953.34 & $1.50\times10^{-15}$ & Parkes & 4.07$^{\circ}$ & 0.2$^{\circ}$ & 2.76\\
	\hline
  \end{tabular}
  \label{tab:1}
  \end{center}
\end{table*}

%
\acknowledgments
{\bf Acknowledgments} We are grateful for stimulating discussions and the high-quality data set from the IceCube Collaboration.  We appreciate helpful suggestions from an anonymous referee.
\newpage

\bibliographystyle{aasjournal}
\bibliography{references}


\end{document}